# Applying a System Dynamics Approach for the Pharmaceutical Industry: Simulation and Optimization of the Quality Control Process

EVRIPIDIS P. KECHAGIAS,
School of Mechanical Engineering
National Technical University of Athens
Zografou 15780
GREECE

DIMITRIOS M. MILOULIS
School of Economic Sciences
National and Kapodistrian University of Athens
Athens 10559
GREECE

GEORGIOS CHATZISTELIOS
School of Mechanical Engineering
National Technical University of Athens
Zografou 15780
GREECE

SOTIRIS P. GAYIALIS
School of Mechanical Engineering
National Technical University of Athens
Zografou 15780
GREECE

GEORGIOS A. PAPADOPOULOS
School of Mechanical Engineering
National Technical University of Athens
Zografou 15780
GREECE

*Abstract:* As countries interact more and more, technology gains a decisive role in facilitating today's increased need for interconnection. At the same time, systems, becoming more advanced as technology progresses, feed each other and can produce highly complex and unpredictable results. However, with this ever-increasing need for interconnected operations, complex problems arise that need to be effectively tackled. This need extends far beyond the scientific and mechanical fields, covering every aspect of life. Systemic Thinking Philosophy and the System Dynamics methodology now seem to be more relevant than ever and their practical implementation in real-life industrial cases has started to become a trend. Companies that decide to implement such approaches can achieve significant improvements to the effectiveness of their operations and gain a competitive advantage. This research, influenced by the Systemic Thinking Philosophy, applies a System Dynamics approach in practice by improving the quality control process of a pharmaceutical company. The process is modeled, simulated, analyzed, and improvements are performed to achieve more effective and efficient operations. The results show that all these steps led to a successful identification and optimization of the critical factors, and a significant process improvement was achieved.

*Key-Words:* Process Modeling, Simulation, Optimization, Quality Control, Pharmaceutical, System, Dynamic

## 1 Introduction

There is no doubt that technology is ever-increasingly affecting industrial operations and everyday life. Complex systems are constantly being developed in order to facilitate operations. A system is a substance with a set of elements called "entities" that have the characteristic that each one of them interacts or is associated with at least one other of the same substance. An entity that is not associated with or does not interact with any element of a system is not part of the system.

A subset of the elements of a system is a system itself and is called a sub-system. The cognitive field of interdisciplinary systems study is called (general) systems theory, or systemic science. Systemic science investigates the organization and abstract properties of matter and intellect in order to discover general principles that govern various concepts regardless of a specific conceptual framework to which they belong, their essence, their type or their spatial/temporal scale of existence.

Barry Richmond, a well-known pioneer of systemic thinking and system dynamics, was the first to formulate, back in 1987, the concept of systemic thinking. According to him: "As interdependence increases, we must learn to acquire knowledge in a new way. It is not enough to try to become better only in our field. We need to have a common language and framework to share our expertise, expertise and experience with "local experts" from other parts of the internet. Only then will we be ready to act responsibly" [1].

It becomes clear that today's need for interconnected operations requires applying systemic thinking in practice. After all, systemic thinking is the tool that can enable us to deal with all the particularly complex problems that arise. Many researchers and scientists in the field of systemic thinking agree with Richmond and confirm the contribution of systemic thinking to addressing these complex problems [2, 3, 4, 5].

Since Richmond's initial formulation of the term "systemic thinking" in 1987, various attempts have been made to explain it and prove its importance. Among others, attempts have been made to answer questions such as: "What makes systemic thinking so difficult to define? Why is it constantly being redefined? What is missing each time?" According to the most popular definitions, A system can be considered as a set of parts that often interact and are interdependent with each other, which, when combined, form a unified entity. Based on the aforementioned view on what is considered a system, it becomes clear that "systemic thinking" could also be defined as a type of system [1].

Systemic thinking is literally a system of thinking about systems. As with most systems, systemic thinking consists of three types: elements (in this case, attributes), interfaces (how these attributes relate to and/or feed back together), and a function or purpose. In particular, the less obvious part of the system, its function or purpose, is often the most critical determinant of systemic behavior [2].

Contrary to the fact that not all systems have an obvious goal or purpose, systemic thinking always has a specific purpose and goal(s). In order to explain its definition, especially to those unfamiliar with the concept, it is essential to disclose this purpose. Therefore, a prerequisite for a correct definition of systemic thinking must be to treat it as a goal-oriented system. To achieve this, the definition must include all three of the above types of operations (elements, interfaces, and objectives).

Barry Richmond, the pioneer of systemic thinking, defines systemic thinking as "the art and science of drawing reliable conclusions about behavior through an ever-deeper understanding of the underlying structure" [1]. As he believes, systemic thinking enables us to see both the small and the big picture [1]. Based on this definition as well as various other approaches identified in the literature [1, 6, 7, 8, 9, 10], we can conclude that systemic thinking allows us to:

- Recognize interfaces: This is the basis of systemic thinking. This ability involves recognizing key connections between components of a system.
- Identify and understand the cause-effect loop: Systemic behavior is strongly affected by the cause-effect loop created by various systemic activities. Therefore, it is essential to understand the results of one process and evaluate them as to how they affect related processes.
- Understand the systemic behavior: It is essential to recognize that each part of a system has a significant role to play. Therefore, the interconnection and feedback between the various parts of a system becomes essential to understand the systemic behavior.
- Differentiate the types of accumulations, flows, variables: Accumulations are the resource groups found inside a system. Flows represent the modifications, increases or decreases, made to the accumulations. Variables are the elements that can be modified in order to affect accumulations and flows.
- Identify and understand non-linear effects: This is related to non-linear accumulations and flows related to linear variables. To avoid confusion, non-linear relationships are separated.

- Understand dynamic behavior: Interfaces, how they are combined in feedback loops, and how feedback loops affect and consist of aggregates, flows, and variables create dynamic behavior within a system. This behavior is difficult to understand without training. Differentiating the types of aggregates flows, and variables, as well as recognizing and understanding non-linear relationships, are both keys to understanding dynamic behavior.
- Reduce the complex systemic behavior: Systemic thinking allows us to use different model views and tools in order to create the system's models. This ability allows us to model even very complex systems by breaking them into simpler and easier to understand sub-systems and manage them effectively. This simplification process aims to find the complex parts of the system and break them into simpler parts until they can be easily managed. Also, this process may include the elimination of information that cause unneeded complexity without offering any significant advantage for the system's operation.
- Recognize and understand various systemic levels: This ability is similar to Barry Richmond's thinking about the small and the big picture. It includes the recognition and understanding of different levels of systemic complexity. Thus, this means being able to recognize both the system as a whole as well as each one of the sub-systems.

Dynamic system evolution essentially represents the new states of a system that arise from the currently studied one as well as their differentiation from it. In most cases, this representation follows a deterministic behavior, in other words, the state of a dynamic system unambiguously defines its evolution in the space of situations. There are, however, some cases where unpredicted facts influence the representation leading to a level of variability and presenting a stochastic behavior.

Dynamic system modeling is a problem-solving approach, through computer simulation, based on systemic thinking and the concept of system feedback, aiming to provide a clear picture of the system's actual behaviour [11]. Forrester firstly presented such a framework in 1956 for presenting and managing economic operations in a systemic way [12].

The philosophy of system dynamics is that complex information interconnections exist in all systemic operations and that all of these operations contain some degree of uncertainty and variability in a non-linear and non-directly comprehensible way [13].

Apart from the financial sector, dynamic systems modeling has already been widely applied in the business sector, especially in business research problems [14, 15, 16, 17]. The primary purpose of these models is to provide an overview of the non-linear behavior of the system, with the ultimate goal of facilitating decision-making by stakeholders and business policy makers.

VENSIM, Ventana's simulation environment, is such a "system dynamics" modeling software for developing, exploring, simulating, analyzing, and optimizing all created models. This tool was specifically developed aiming to improve the effectiveness of process reengineering and can offer a significant aid when trying to optimize operations. Utilizing its ability to disseminate information in an understandable way has set it as one of the primary tools used for learning purposes. This is why in this research, we decided to utilize the VENSIM tool for the process simulation [18].

More specifically, our research, influenced by systemic thinking and system dynamics, tries to improve the quality control process in the pharmaceutical industry. To the best of our knowledge, systemic thinking applications are scarce in research papers and most notably in pharmaceutical quality control, where we believe our research is the first one to be conducted.

In the following sections, after conducting research on today's best practices and trends for the pharmaceutical sector, we then utilize this knowledge and combine it with dynamic systemic thinking. More specifically, we simulate the quality control process in the pharmaceutical industry and, utilizing the VENSIM System Dynamics method, we experiment by changing various factors in order to achieve an improved, near-optimal process.

## 2 Best Practices for the Pharmaceutical Industry

For this research, it was deemed appropriate to study and present research on the best practices and trends applied internationally in the pharmaceutical industry while also mentioning specific examples of successful implementations.

### 2.1 Implementing the Lean Philosophy

It is a fact that the pharmaceutical industry is associated with complex service systems that deal with different types of clients (patients, suppliers, distributors, hospital staff), use a wide range of staff with different role combinations and perform a variety of critical procedures. They, therefore, need to incorporate advanced technological solutions to improve the accuracy and speed of their services.

Numerous research publications concern the implementation of lean processes for their simplification, saving time and costs and in general increasing their efficiency and effectiveness. The methodology of lean processes comes from the field of production. Precisely, the Japanese company Toyota, first-applied in its production the lean procedures. The methodology focuses on problem-solving, collaboration, simplification and ultimately process improvement. The philosophy of this lean approach also functions as an effective model for the pharmaceutical industry. Studies have shown that this methodology can be effectively applied to a hospital pharmacy's full range of procedures [19].

The simple principles, which include "Muda" and "Kaizen", concern the continuous identification and elimination of useless procedures, documents, steps, etc., in order to ultimately leave only those that have some sort of added value for the recipient of the services. "Muda" or "waste" means "any human activity that absorbs resources but does not create value". "Kaizen" can be considered as the effort to continuously improve, as fast as possible, while operating in an ever-changing, demanding workplace. Both "Kaizen" and "Muda" are related to the ability of a company to avoid useless processes, continuously improve, adopt applicable workflow models and develop leaders acting according to teamwork philosophy and respect [20].

The University of Minnesota Medical Center (Fairmount) is a typical successful example of an organization that implemented the lean philosophy to improve the operational workflow, reduce "Muda", and eliminate unnecessary expenses. By applying lean techniques, pharmaceutical waste was reduced by 40%. Reducing the waste led to annual savings of $ 275,500. At the same time, there were achieved fewer shortages of medicines, and expired products were reduced (20%). It is estimated that about $ 50,000 were saved due to the improvements in inventories [21].

Another successful example of the application of lean procedures is the outpatient pharmacy at Yale-New Haven Hospital, in order to perform all necessary changes to its operations for avoiding waste and unneeded expenses. Applying the lean philosophy tools, the team of employees in the pharmacy carried out an analysis of the way of working, mapped the workflow and performed an impact analysis. Thirty-eight opportunities were identified to reduce waste and increase efficiency. Three of the pharmacy procedures (prescription testing, product verification and delivery to clinics) accounted for 24 of the 38 opportunities. Applying the methodology of lean processes, 6 out of the total 20 processes were deemed unnecessary and were removed as they offered no value for the customers. The check process of the recipes was significantly shortened (33%), as was the product verification process (52%). At the same time, the delays for product deliveries were widely reduced (47%) [22].

## 2.2 Other Trends for Pharmaceutical Operations Improvement

Another technique that has been applied in many pharmaceutical industries concerns techniques for identifying and controlling the risks associated with their operation. Risk can be considered as the likelihood of an occasion and the results arising from it. As with all occasions, the risk can become both an opening for gaining a competitive advantage as well as dangers that may cause serious issues. Both of these viewpoints of risk require assessment in order to be either be exploited or discarded. Risk assessment techniques are widely used in the dangerous nuclear, oil and gas industries. A key element of these techniques is that they involve identifying risk mitigation actions and evaluating their effectiveness. However, these techniques are relatively uncommon in the clinical pharmaceutical industry. Of course, if one considers the plethora and impact of risks in the pharmaceutical industry, then the need for risk assessment techniques becomes evident [23].

In addition, a practice that has been widely used in the pharmaceutical industry to improve its performance involves simulating its operation using specialized software [24, 25]. In the pharmaceutical industry, simulation can have many benefits. It can be used to enable making decisions under uncertainty, such as calculating the gains and losses from the implementation of a new drug lending system in clinics. It can also be used to conduct comparative studies where, by running different scenarios using different resources, the optimal way of conducting a process can finally be identified. The simulation can generally be applied to the entire range of the pharmaceutical industry's procedures, contributing significantly to its more efficient and effective operation [26].

Finally, a prevailing direction found in the literature concerns techniques for automating pharmaceutical procedures. The automation of the procedures is done with the use of technology, helping the industry achieve cost reduction while improving the provided services. Process automation is utterly different from using robots to make products. This use is called industrial automation and aims to replace natural human work such as building a car using robotic assistants. In contrast to industrial automation, process automation is interpreted as a way in which software and various applications are used to automate repetitive business processes [27, 28, 29, 30].

A typical example of application in the pharmaceutical industry is the use of an RFID system to automate its various procedures. An RFID system consists of labels or tags and RFID reading devices that communicate dynamically. This system operates by moving the RFID tag close to the reading device that then gets activated and reads all data related to a specific tag. In order for an RFID system to operate effectively, there is also a need for software that translates and transfers all related information. Such systems could be used to order and receive medicines, check prescriptions, and enter data into information systems [31].

The aforementioned methods and techniques are just some examples of the potential for improving the efficiency of the pharmaceutical industry. In fact, in order to achieve the best possible results, before applying any method, a team of experts needs to conduct a thorough study in order to identify the appropriate methods or the combination of methods that need to be applied in each individual case. Given the conditions in the pharmaceutical sector and in order for this industry to cope with the ever-increasing challenges related to the provision of its services, today more than ever, it is considered necessary to try to optimize its processes.

## 3 Using System Dynamics for Improving the Quality Control Process of a Pharmaceutical Company

For this research, we decided to apply systemic thinking focusing on the quality control process of a pharmaceutical company. Therefore, utilizing the VENSIM System Dynamics method, we simulate the process and modify various parameters aiming at optimizing the performance.

### 3.1 The Company's Need for Improvement

For this study, data were obtained from a pharmaceutical company and specifically from the quality control department of pharmaceutical and parapharmaceutical products. The production process is complicated and demanding as from the total number of produced products, only a relatively small percentage comply with the strict disposal regulations set by law and may be allowed for sale.

Possible disposal of unsuitable products may have severe consequences for consumers' health but of course, also a huge financial disaster for the company itself. Therefore, proper quality control and practical training of testers is required in order to avoid this catastrophic consequence. The simulated and improved system, in this research, is the process of training auditors, ultimately aiming at the most effective product quality control.

The executives of the pharmaceutical company are often concerned about the "image" that the company has outwards, and mainly about issues related to its reliability and the quality of its products, as a result of which they periodically ask their customers to evaluate the quality of the products in order to identify weak points. Often their comments are very positive about the quality of their products, but not so for the speed and delivery of orders. Due to the fact that this is a pharmaceutical company, it is required to deliver the products on time without any delays, as these delays create great dissatisfaction among customers, and the company itself may receive negative reviews. Of course, when the customers cannot receive their orders on time, they show their intense dissatisfaction towards the company for its non-observance of its obligations. Finally, in some cases, the products are returned to the pharmaceutical company as customers report damaged packaging and/or side effects.

The management of the pharmaceutical company is now more and more interested in the company's image since there are quite a few complaints about delays and product returns. For this reason, they hire extra staff so that the quality control of the products becomes even more meticulous.

Two key factors determine the decision to hire staff to handle quality controls:
- the number of complaints made by customers
- the number of employees at any given time

The staff hired to carry out the quality control should first be trained for a period of a few months before fully understanding the complex procedures of the pharmaceutical company. The newly trained testers do not immediately check the products that are ready to be given to customers, as in this case, the probability of an error is increased. The new testers are always trained by experienced staff. Training an employee is quite a time-consuming process as an experienced employee who has a reasonably large workload will have to devote about 50% of his time to train someone else properly. The policy of the pharmaceutical company is against the dismissal of the people who carry out the tests, but gradually the number of testers is decreasing. After completing their training, the testers remain in the pharmaceutical company for an average of three years. The number of testers employed by the pharmaceutical company depends directly on its production, as the higher the output, the more testers will have to work. Also, the testing time devoted to each product depends on the production levels in each time period. The unit of time used for this study is "months".

## 3.2 System Modeling and Data & Correlation Analysis

According to Figure 1 presenting the Stock Flow diagram, there are only two accumulations (Stocks), i.e., elements of the system where something is "accumulated", the Trainee Testers and Trained Testers. At the same time, all the other variables are either fixed (their names are written with uppercase letters) or are auxiliary variables.

Finally, in Figure 1, 3 flows can be observed:
- the hiring rate that fills the Trainee Testers accumulation
- the training completion rate that empties the Trainee Testers accumulation and fills the Trained Testers accumulation
- the quitting rate that empties accumulation Trained Testers

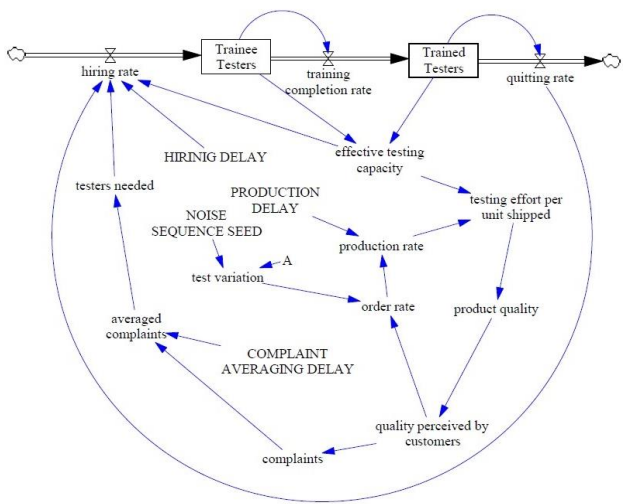

Fig. 1: Stock-Flow Diagram

The equations that characterize the variables of the model are presented in Table 1. These equations are essential in order to perform the model's simulation and followingly, make all necessary improvements.

Table 1. System's Equations

(1) A=0
(2) averaged complaints = SMOOTH(complaints, COMPLAINT AVERAGING DELAY)
(3) COMPLAINT AVERAGING DELAY = 2
(4) complaints = (3 / quality perceived by customers) -2
(5) effective testing capacity = Trained Testers - 0.5*Trainee Testers
(6) FINAL TIME = 120
(7) hiring rate = MAX(0, quitting rate + (testers needed – effective testing capacity) / HIRING DELAY)
(8) HIRING DELAY = 2
(9) INITIAL TIME = 0
(10) NOISE SEQUENCE SEED = 958
(11) order rate – 10000*quality perceived by customers * (1 + test variation)
(12) product quality = IF THEN ELSE (testing effort per unit shipped <0.01, 100* testing effort per unit shipped, 1+ 10*(testing effort per unit shipped – 0.01))
(13) PRODUCTION DELAY = 3
(14) production rate = DELAY FIXED (order rate, PRODUCTION DELAY, order rate)
(15) quality perceived by customers = SMOOTHI (product quality, 6, 1)
(16) quitting rate = Trained Testers / 36
(17) SAVEPER=0.125
(18) test variation = STEP (0.2, 5) * ((1-A) +A*RANDOM UNIFORM(-0.5, 0.5, NOISE SEQUENCE SEED))
(19) testers needed = IF THEN ELSE (averaged complaints<0.5, 0, 200*(averaged complaints -0.5))
(20) testing effort per unit shipped = effective testing capacity / production rate
(21) TIME STEP =0.125
(22) Trained Testers = INTEG (training completion rate - quitting rate, 100*24/23)
(23) Trainee Testers =INTEG (hiring rate – training completion rate, (3/36)*(100*24/23))
(24) training completion rate = Trainee Testers / 3

At this point, we will try to explain how the above equations were chosen and the functionality that they serve for the correct simulation of the process. The data for the modeling of the system were collected thanks to the valuable help of the pharmaceutical company.

The simulation will start at time zero and will be completed after 120 months, as shown in equations (6), (9). It is known that a process is in equilibrium when the contained variables do not change over time and remain constant. In our case on the one hand, we have the new testers (Trainee Testers) who are trained for a period of about three months (their training process is expressed through the variable "training completion rate"), and on the other hand, we have the testers who have passed the basic training stage (Trained Testers), whose experience gradually increases and they stay in the pharmaceutical company for an average period of three years (36 months). The pharmaceutical company periodically hires new staff (the flow variable hiring rate is used to express the number of recruitments made by the pharmaceutical company) either to fill the gaps created by the departure of employees (quitting rate) or because production has increased, resulting in the need for more auditors to inspect products without much delay.

Therefore, for the process to be in equilibrium, the variables "Trainee Testers" and "Trained Testers" should not change, and for this reason, we should pay close attention to their initial values. In addition, the flow variable "hiring rate" must be equal to the flow variable "training completion rate", and respectively the flow variable "training completion rate" must be equal to the flow variable "quitting rate". We should note that the unit of measurement for all of the aforementioned variables is common (number of employees).

Equation (24) expresses the value of the variable "training completion rate" equal to "Trainee Testers / 3". This variable, as we see, represents the rate at which the training of auditors is completed (their training lasts about three months, hence the result and the denominator of the fraction of equation (24)). Something similar happens with the variable "quitting rate" in equation (16) which expresses the rate at which employees leave the company after a period of about 36 months. The equation at this point is easy to understand and no further explanation is required. These two equations should be utilized to ensure the stability of the model. We previously mentioned that "training completion = quitting rate" being an equilibrium condition for the "Trainee Testers variable". This means that "Trainee testers / 3 = Trained testers / 36" and the initial value for the stock variable "Trainee Testers" is obtained from solving this system.

To avoid having to do such complex calculations by hand, we use the Equation (22) of "Trained Testers". The relation comes from the equation "Trained Testers = INTEG (training completion rate - quitting rate, Initial Value of Trained Testers)", where the initial value of the variable "Trained Testers" is "100 * 24/23". Having determined the value of the variable "Trained Testers", the equation (23) for "Trainee Testers" is easily obtained, i.e., "Trainee Testers = INTEG (hiring rate - training completion rate, Trained Testers * (TRAINING PERIOD / EMPLOYMENT PERIOD))", where "TRAINING PERIOD / EMPLOYMENT PER = 3/36" and "100 * 24/23" is the initial value of the "Trained Testers" variable.

The auxiliary variable "effective testing capacity", presented in equation (5), expresses the actual number of controllers involved in controlling the products produced. According to the quality manager of the pharmaceutical company, an experienced tester should devote 50% of his working time to the training of a new tester, which has the effect of reducing the number of employees involved in conducting tests. Initially, there are a number of testers working in production, and some of them (this number is equal to the number of the trained trainees) will have to devote 50% of their time to the training of new employees. The constant "PRODUCTION DELAY", as seen in equation (13), expresses the delay that exists for the production of the products that the customers have ordered. While the constant "HIRING DELAY", as presented in equation (8), expresses the time required to hire new employees from the moment the pharmaceutical company decides to do so.

The flow variable "hiring rate" models the way the company decides to recruit staff in order to train them for the quality testing. The value of the variable is a consequence of a recommended action of two factors:
- the number of testers leaving the company and
- the requirements arising from the production

There is, of course the possibility that no intake is required. For this reason, in equation (7), the "MAX" function is used, which essentially compares the two arguments of the function. This also avoids the negative values that the variable can possibly receive in case the available testers are more than those that are really needed at the given time. Equation (20) shows the relation that represents the value of the variable "testing effort per unit shipped" corresponding to the time (for testing) that is dedicated separately for each unit of product produced by the factory. This depends on the pace of production and the number of testers who can handle product testing.

The more time devoted to inspecting the products produced, the less likely it is that a manufacturing error will be detected. The quality of the products that customers will receive depends solely on how intensive and careful the tests are performed. This fact is accurately reflected in equation (12), which expresses the quality of the product. The variable "product quality", as seen in Figure 2, is essentially a scale of product evaluation that has as its sole criterion the level of tests performed. The IF THEN ELSE function is essentially used to implement the scale on which the quality of the final product is graded. If the performed test is satisfactory, then a score greater than "1" is obtained else, the score is less than "1". This method is popular in large companies that generate scales of ratings on their own using specific criteria so that the results are easily manageable.

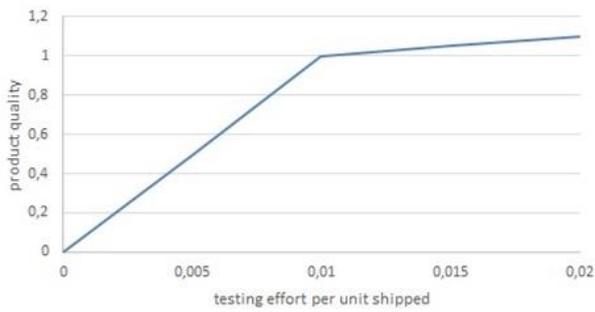

Fig. 2: Product Quality in Relation to Testing Effort

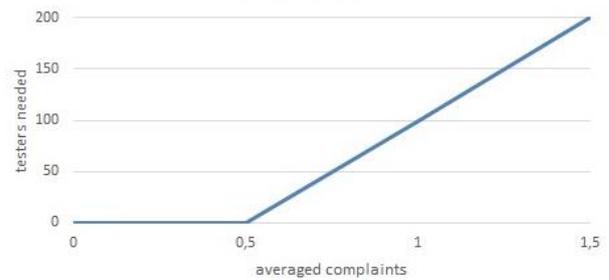

Fig. 3: Testers Needed in Relation to the Average Number of Complaints

Equation (15) shows the relationship of the variable "quality perceived by customers" that actually expresses the opinion that customers have about the quality of the product they receive. When the variable "quality perceived by customers" gets the value "1", it means that only one complaint arrives at the pharmaceutical company per month. We see, according to Figure 1, that the variables that determine the quality as perceived by the patient are the "order rate", "production rate", "testing effort per unit shipped", and "product quality". We find the existence of a loop with the result that in the long run, this variable affects itself. In this case, we are dealing with an evaluation scale where the only one who gives scores is the customer. The better the customer's opinion about the quality, the larger the orders he will request. Because of this loop, it is impossible to determine the initial values of the variables involved, and for this reason, we use the "SMOOTH I" function, which allows us to determine an initial value for the output of the function. With equation (15), we determine that the initial output for the "SMOOTH" function is equal to "1". Then VENSIM can use this value so that "quality perceived by customers" is equal to "1". From there, other variables' values can be easily calculated.

The "complaints" variable, presented in equation (4), is an indicator of the number of complaints customers have made. We use the word "indicator" because it is impossible to quantify the complaints a customer may have. But what we can do is use the data we have and create rating scales like we did before. This will help us a lot in creating a model of high accuracy, while at the same time, these scores are easy to understand and manage. Also, the constant "COMPLAINT AVERAGING DELAY" of equation (3) expresses the interval between the receipt of the products until the customers make the first complaints. The number of new testers as seen in Figure 3, is based on an "IF THEN ELSE" function, determined by the number of complaints (average complaints) expressed by customers.

As we have already seen, "IF THEN ELSE" functions we used in two cases. First to categorize the quality of the final product (product quality) depending on the intensity of the tests devoted to each unit (Figure 2) and then to correlate the number of testers that the company should hire based on the number of complaints expressed by the customers (Figure 3). In the first diagram, we find that the more time we devote to the testing of each product, the better the quality of the final product becomes. However, the increase is not always in the exact same way, as from one point onwards even though we devote enough time for quality control of each unit separately this does not imply a significant improvement in overall quality.

In Figure 3, we find that the higher the index representing the number of complaints made by customers, the more testers are needed. Of course, there is a limit, as if the complaints made are limited, the pharmaceutical company is not obliged to hire new staff. After all, an increase in the workforce is not a solution to all the problems as there will always be some defects in production. At the same time, hiring staff implies an increase in expenses, and it is possible that the additional costs that will arise may not be easily depreciated.

Also interesting is the way in which orders are modeled, as shown in equation (11) hat expresses the "order rate". This depends on two factors, first of all on the opinion and how satisfied the customers are with the pharmaceutical company, which is reflected in the opinion they have about the quality of the products they receive. Secondly, to make the model even more realistic, we have introduced the element of luck. The course of sales is not constant but can be determined by random events that we cannot take into account in advance, which is why in equation (18), we use the function "RANDOM UNIFORM", a function that produces random numbers, and the constant "SEQUENCE NOISE SEED" is used to initialize the function generator that generates the random numbers, so the value of "958" does not matter. At the same time, we identify the existence of the constant "A" that is used to determine the price that the orders will receive. When the constant "A" has the value "0", then the value of the "test variation" variable follows a stepwise path where after five months, the value instantly becomes "0.2". When "A = 1", then the test variation takes random values from "-0.1" to "+0.1". Finally, at this point, we should note that if the orders are fixed, then all the other variables of our model will remain constant.

The factory production rate (14) depends on two factors, the "order rate" and the delay that always exists for the adjustment of the production line according to the existing needs. The production varies according to the needs, and if many large orders suddenly arise, then the company will have to procure the appropriate raw materials, buy additional staff, etc. and obviously, to do all this takes some time. The company produces the quantities requested by the customers, and for this very reason, the "DELAY FIXED" function is used to indicate that the quantities of product produced in the company are the same as those requested from time to time by the customers, except that orders production preparation takes three weeks.

### 3.3 Simulation

Having completed the analysis of the equations, we are ready to simulate the model we created. For this purpose, we execute the "SyntheSim" command, and the model of Figure 4 is displayed with the various sliders of the parameters that we can modify.

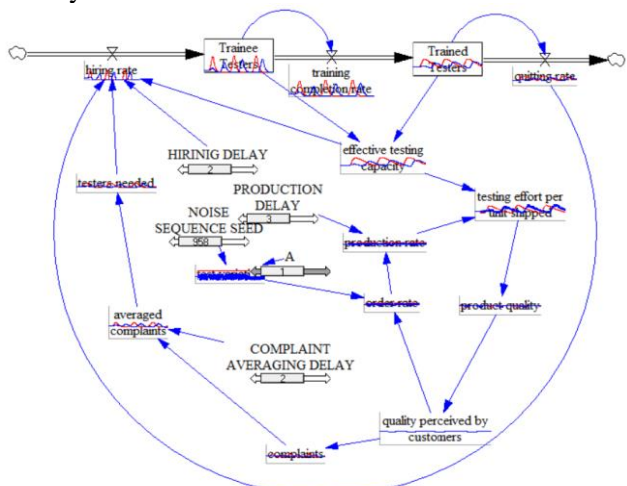

Fig. 4: Existing SyntheSim Model

Figures 5,6, and 7 show the graphs of Causes Strips for the variables that we consider essential, i.e. Trainee Testers, Production Rate & Trained Testers.

Of particular interest is Figure 6, which also graphically confirms what we had already predicted. The course of orders and production is precisely the same with the only difference that in terms of production (production rate), there is a slight delay. In the other graphs, the course of the values of the variables is recorded. All these elements are very important, and the fact that they are represented graphically reflects the necessity of their use. Utilizing the right ones can be a very useful tool to make the most of the right decisions for improving the process.

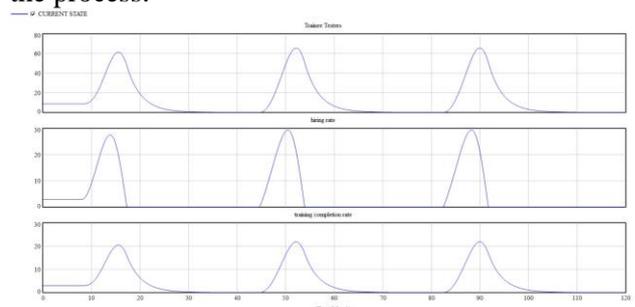

Fig. 5: Simulation Results for the Variables "Trainee Testers", "Hiring Rate", "Training Completion Rate" (Unit of Measure for all Three Variables is the Number of Employees)

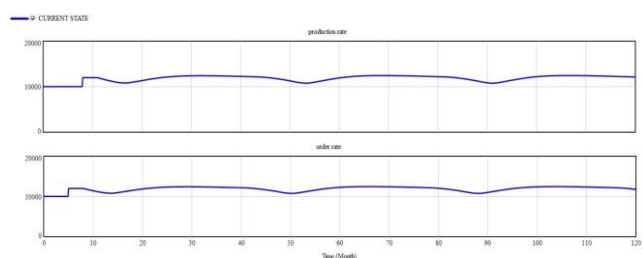

Fig. 6: Simulation Results for the "Production Rate" and "Order Rate" Variables (Unit of Measurement for Both Variables are the Product Units)

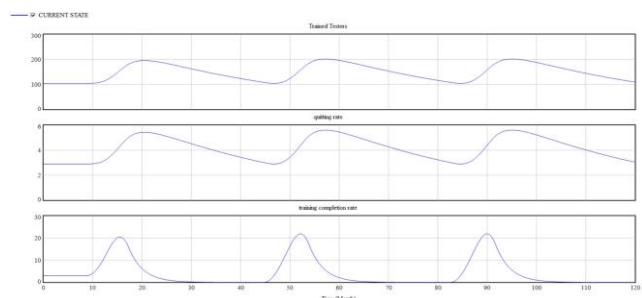

Fig. 7: Simulation Results for the Variables "Trained Testers", "Quitting Rate", "Training Completion Rate" (Unit of Measurement for all Three Variables is the Number of Employees)

Figures 8 and 9 present the Causes Tree & Uses Tree diagrams for the essential variables of the current situation.

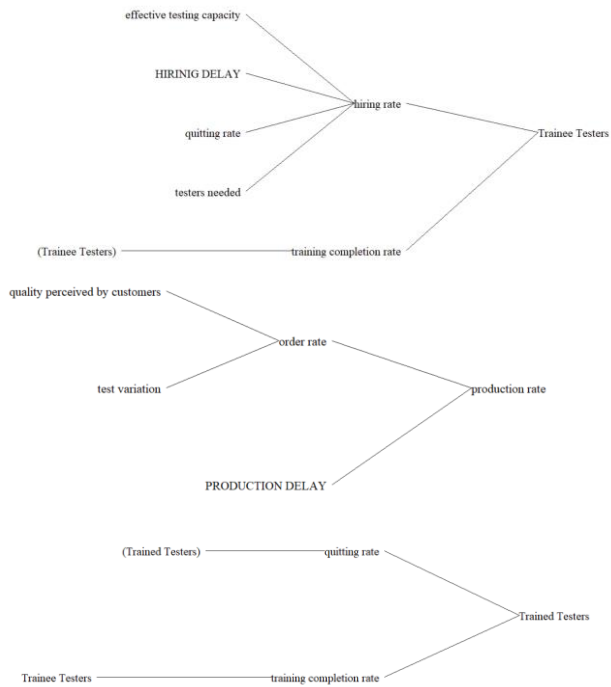

Fig. 8: Causes Trees (Current Situation)

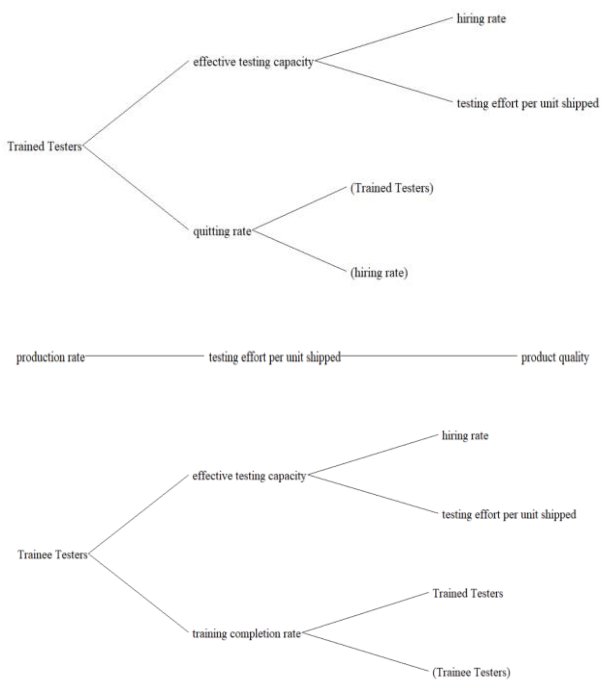

Fig. 9: Uses Trees (Current Situation)

### 3.4 Model Performance Improvement & Evaluation

The System Dynamics methodology at a first level allows us to create a model that describes a process. Then we can modify the model and show the impact of any changes we bring about. Therefore, in this model, we try to find ways that will improve the performance of our system. The changes we will make are aiming to improve the process by reducing customer complaints and also increasing sales. One way to improve the process could be to reduce the delays in the current model. Specifically, we refer to the constants "HIRING DELAY", "PRODUCTION DELAY" and "COMPLAINT AVERAGING DELAY". However, in order to reduce the delays, changes must be made in the whole process. Another way could be to change the policy of hiring testers. In our suggested solution, we have chosen the second case, i.e. the change in the way of hiring the testers, as seen in Figure 10. The equations that characterize the variables of the newly modified model are presented in Table 2.

Table 2. System's Modified Equations

(25) A=0
(26) average order rate = SMOOTH(order rate, ORDERING AVERAGE PERIOD)
(27) complaints = (3 / quality perceived by customers) -2
(28) effective testing capacity = Trained Testers - 0.5*Trainee Testers
(29) FINAL TIME = 120
(30) hiring rate = MAX(0, quitting rate + (testers needed – effective testing capacity) / HIRING DELAY)
(31) HIRING DELAY = 2
(32) INITIAL TIME = 0
(33) NOISE SEQUENCE SEED = 958
(34) ORDERING AVERAGE PERIOD = 2
(35) order rate = 10000*quality perceived by customers * (1 + test variation)
(36) product quality = IF THEN ELSE (testing effort per unit shipped <0.01, 100* testing effort per unit shipped, 1+ 10*(testing effort per unit shipped – 0.01))
(37) PRODUCTION DELAY = 3
(38) production rate = DELAY FIXED (order rate, PRODUCTION DELAY, order rate)
(39) quality perceived by customers = SMOOTHI (product quality, 6, 1)
(40) quitting rate = Trained Testers / 36
(41) SAVEPER=0.125
(42) test variation = STEP (0.2, 5) * ((1-A) +A*RANDOM UNIFORM(-0.5, 0.5, NOISE SEQUENCE SEED))
(43) testers needed = 0.01*average order rate
(44) testing effort per unit shipped = effective testing capacity / production rate
(45) TIME STEP =0.125
(46) Trained Testers – INTEG (training completion rate - quitting rate, 100*24/23)
(47) Trainee Testers =INTEG (hiring rate – training completion rate, (3/36)*(100*24/23))
(48) training completion rate = Trainee Testers / 3

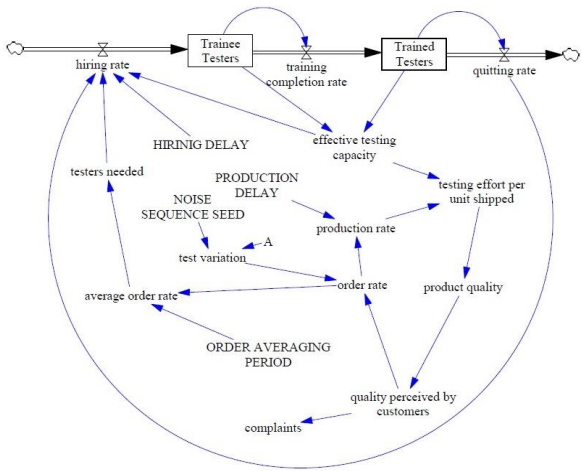

Fig. 10: Proposed Change in Testers Hiring Policy

In the original model, the number of new testers needed to be hired was determined using an "IF THEN ELSE" function whose conditions were determined by the number of complaints, i.e., average complaints expressed by customers. On the contrary, in this new model the number of new testers depends exclusively on the course of the orders or the, i.e., the averaging order rate, and this is clearly reflected in equation (43). Therefore, more importance is given to the speed of production while at the same time the process of hiring testers is simplified, and both factors aim to improve the efficiency of the pharmaceutical company.

Therefore, we see that we are essentially shaping a new policy with such a slight change in the model. To examine the consequences of our changes, it is essential to run the simulation again and compare the results that will appear with the initial ones. Firstly, we execute the SystheSim command as seen in Figure 11, representing the various sliders of the parameters that we can modify.

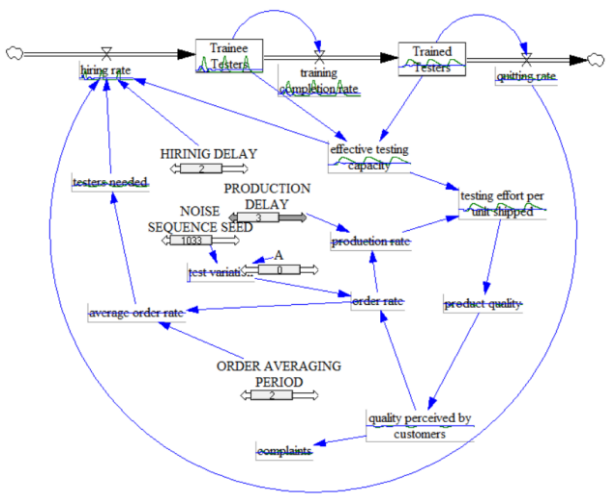

Fig. 11: SyntheSim Model (Improved Situation)

Comparing the results obtained from the Causes Strips for the essential variables based on our changes, i.e., the "Quality Perceived by Customers" and "Trainee Testers", presented in figures 12 and 13, we find that the changes had a significant impact for the process performance. The number of testers to be hired is significantly reduced, while the quality of the products is consistently higher, and the same is true of for what customers think about the pharmaceutical company's services.

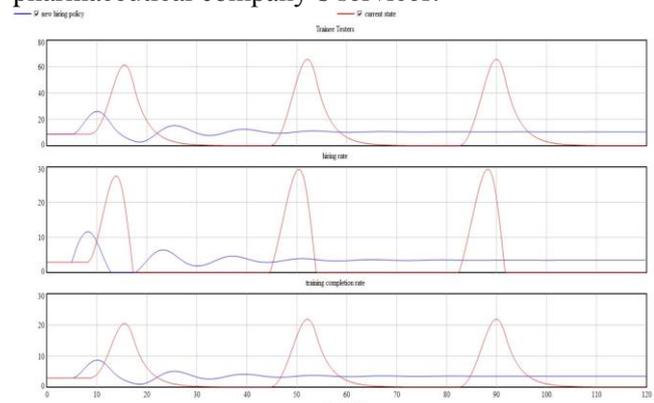

Fig. 12: Comparison of the "Trainee Testers" Needed for Both Situations

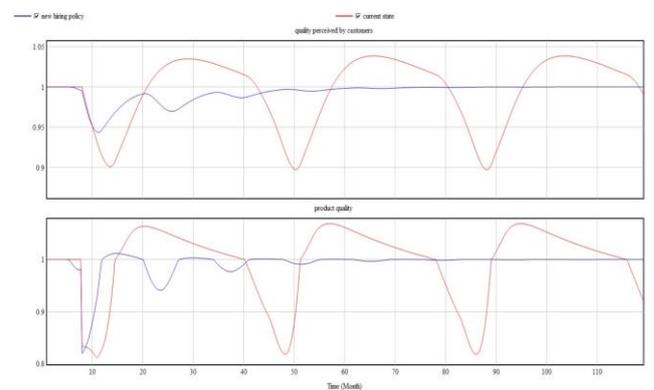

Fig. 13: Comparison of the "Quality Perceived by Customers" for Both Situations

Figures 14 and 15 present the Causes Trees & Uses Trees for the essential variables in the improved situation.

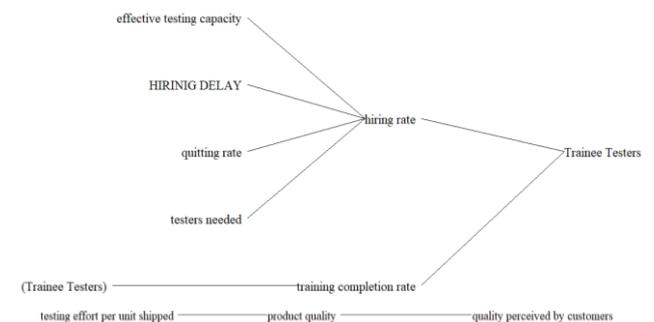

Fig. 14: Causes Trees (Improved Situation)

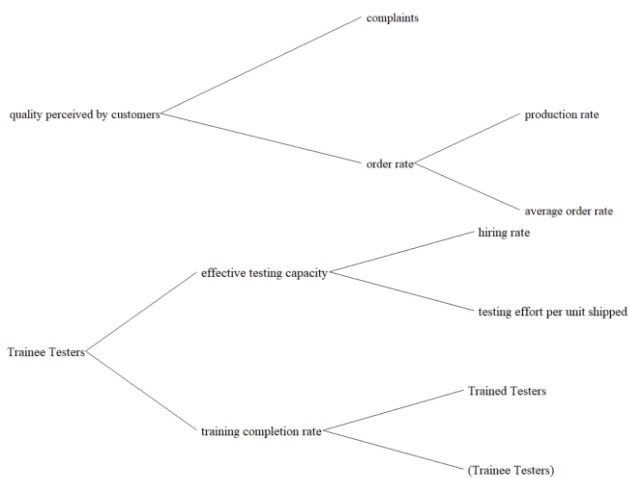

Fig. 15: Uses Trees (Improved Situation)

## 3.5 Additional Possible Improvements

VENSIM enables us to experiment a lot with the model in order to find solutions that will help us improve the performance of the process we have modeled. There are so many trials and changes that can be made to our model, but in this section, we have chosen to consider what will happen if we apply the first proposed solution mentioned in the previous section, i.e., if we reduce the delays of the system. We are essentially changing the constants "HIRING DELAY" and "PRODUCTION DELAY".

Of course, in order to reduce these delays, changes must be made throughout the process. In the future, the pharmaceutical company has the ability to apply them in case the simulation shows encouraging enough results. With the help of graphic representations, we have the ability to see if the changes we will make will have a positive or negative impact. In Figures 16 and 17, we see the changes that occur if we slightly change the constants "HIRING DELAY" and "PRODUCTION DELAY". Increasing the "HIRING DELAY" from 2 to 4 months will lead to a slight reduction in the departure rate but also to a reduction for the needed testers. On the other hand, we see that increasing the "PRODUCTION DELAY" from 3 to 6 months will not have any substantial impact.

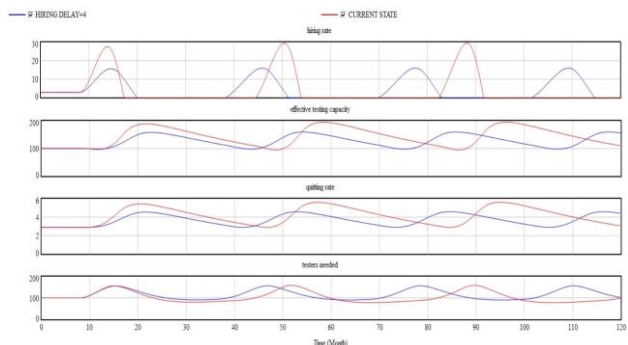

Fig.16: Influence of Change in "HIRING DELAY"

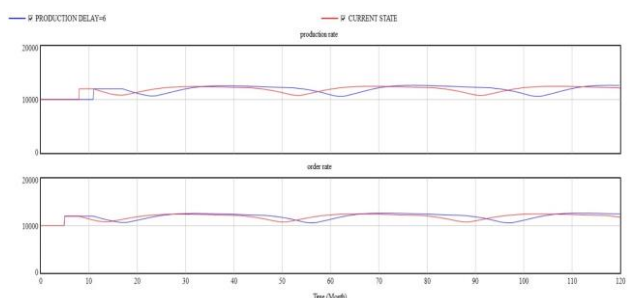

Fig.17: Influence of Change in "PRODUCTION DELAY"

Finally, we decided to study the effect that the change of the constant "A" will have. The value that the constant "A" can get is either "0" or "1". Obviously, the value that the constant "A" will get also determines the shape that the variable "test variation" will have that directly affects the "order rate". If "A = 0", then the variable "test variation" plot looks like a step function. In fact, after five weeks, its value increases instantaneously from the value "0" to the value "0.2". If "A = 1", then the variable "test variation" follows a random and unpredictable course. This is achieved with the help of "RANDOM UNIFORM". "NOISE SEQUENCE" is used to initialize the function generator that produces the random numbers.

As previously stated, orders depend on the opinion of the customers about the pharmaceutical company and on random events that we may not be able to predict. This is precisely what variable "test variation" does by introducing the element of luck into orders. In this way, the company will is able to get a more realistic model by changing the "A" value from "0" to "1".

In Figures 18, 19 we observe the changes resulting from the change of the value of the constant "A". In fact, we find that when "A = 1", better results occur as the new testers to be hired are fewer (testers needed) and the customers are more satisfied with the quality of the products they receive (quality perceived by customers).

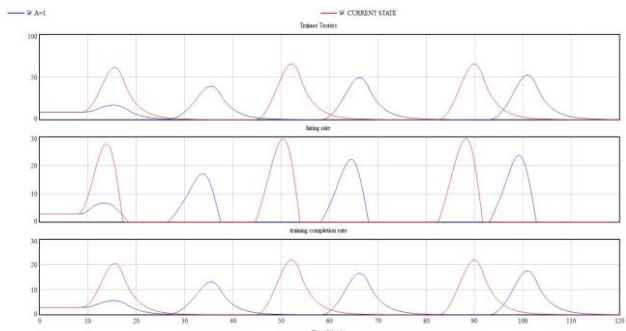

Fig. 18: Influence of Change of Constant "A" to Needed Trainee Testers

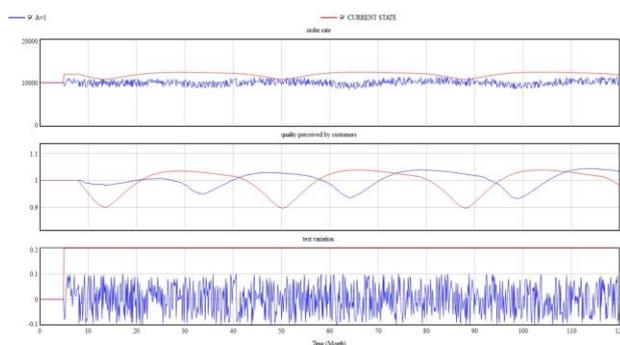

Fig. 19: Influence of Change of Constant "A" to Quality Perceived by Customers.

## 4 Conclusion

It becomes clear that today more than ever, implementing the systemic thinking philosophy in real-life industrial operations is essential for companies to achieve a competitive advantage and operate effectively and efficiently. After all, systems play and will continue to play in the future a vital role for industrial operations. This work reaffirms the above, presenting a real-life implementation and confirming the value of systemic thinking and dynamic systems modeling in industrial operations. The research shows that even small changes in various parameters can have a decisive effect on the model's performance. Utilizing the proposed solutions, the pharmaceutical company can achieve significantly better results in the quality control of the products it produces. In addition, by modifying various other parameters and combining them with the proposed solution, the quality control process can be further enhanced. The contribution of this research is that it sets a path for implementing the systemic thinking philosophy in practice, not only in the quality control process but also at every aspect of the industrial operations. In the future, this research approach can be extended to any industrial operation in order to model, simulate, analyze and optimize it. After all, great potential exists in the industrial sector, so we should make hay while the sun shines.

## Contribution of individual authors to the creation of a scientific article (ghostwriting policy)

## Creative Commons Attribution License 4.0 (Attribution 4.0 International, CC BY 4.0)